\documentclass[11pt]{article}
\usepackage{moriond,epsfig}

\def\be{\begin{equation}}
\def\ee{\end{equation}}
\def\bea{\begin{eqnarray}}
\def\eea{\end{eqnarray}}

\begin{document}
\vspace*{4cm}
\title{HIGH DENSITY QCD PHYSICS WITH HEAVY IONS IN CMS}

\author{Ferenc~Sikl\'er}

\address{KFKI Research Institute for Particle and Nuclear Physics, \\
                    1121 Budapest, Hungary}

\maketitle\abstracts{
The heavy ion program of the CMS experiment will examine the QCD matter under
extreme conditions, through the study of global observables and specific probes.
}

\section{Introduction}

The CMS detector has a large acceptance and hermetic coverage. The various
subdetectors are: a silicon tracker with pixels and strips ($|\eta|<2.4$),
electromagnetic ($|\eta|<3$) and hadronic ($|\eta|<5$) calorimeters, muon
chambers ($|\eta|<2.4$). The acceptance is further extended with forward
detectors ($|\eta|<6.8$). CMS detects leptons and hadrons, both
charged and neutral ones. In the following, capabilities in soft, hard and
forward physics are described. For a very recent extensive review see
Ref.~1.

\section{Soft physics}

\begin{figure}[h]
 \begin{minipage}[c]{0.45\textwidth}
 \includegraphics[width=\textwidth,angle=-90]{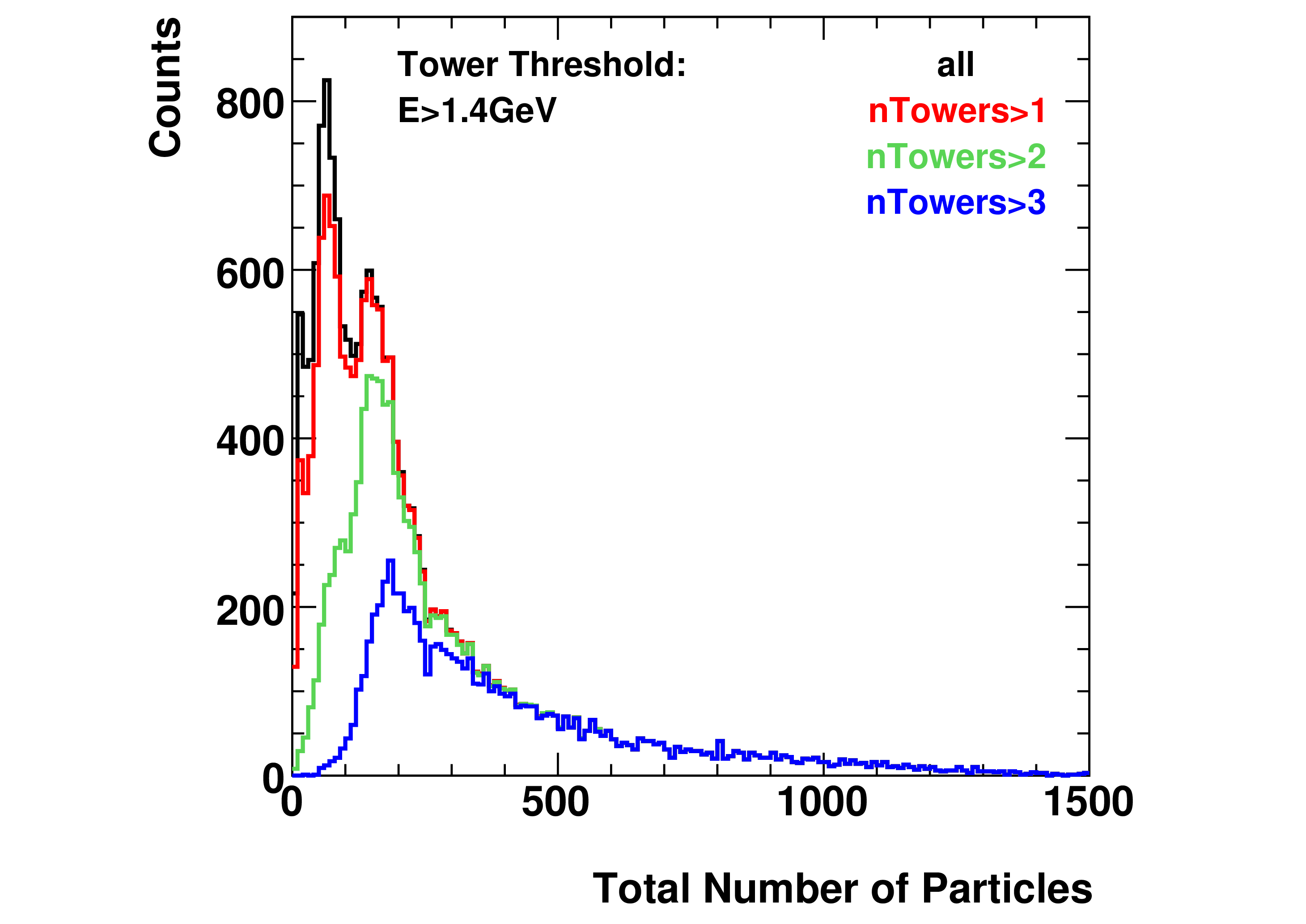}
 \end{minipage}
 \hspace{0.02\textwidth}
 \begin{minipage}[c]{0.45\textwidth}
 \includegraphics[width=\textwidth,angle=-90]{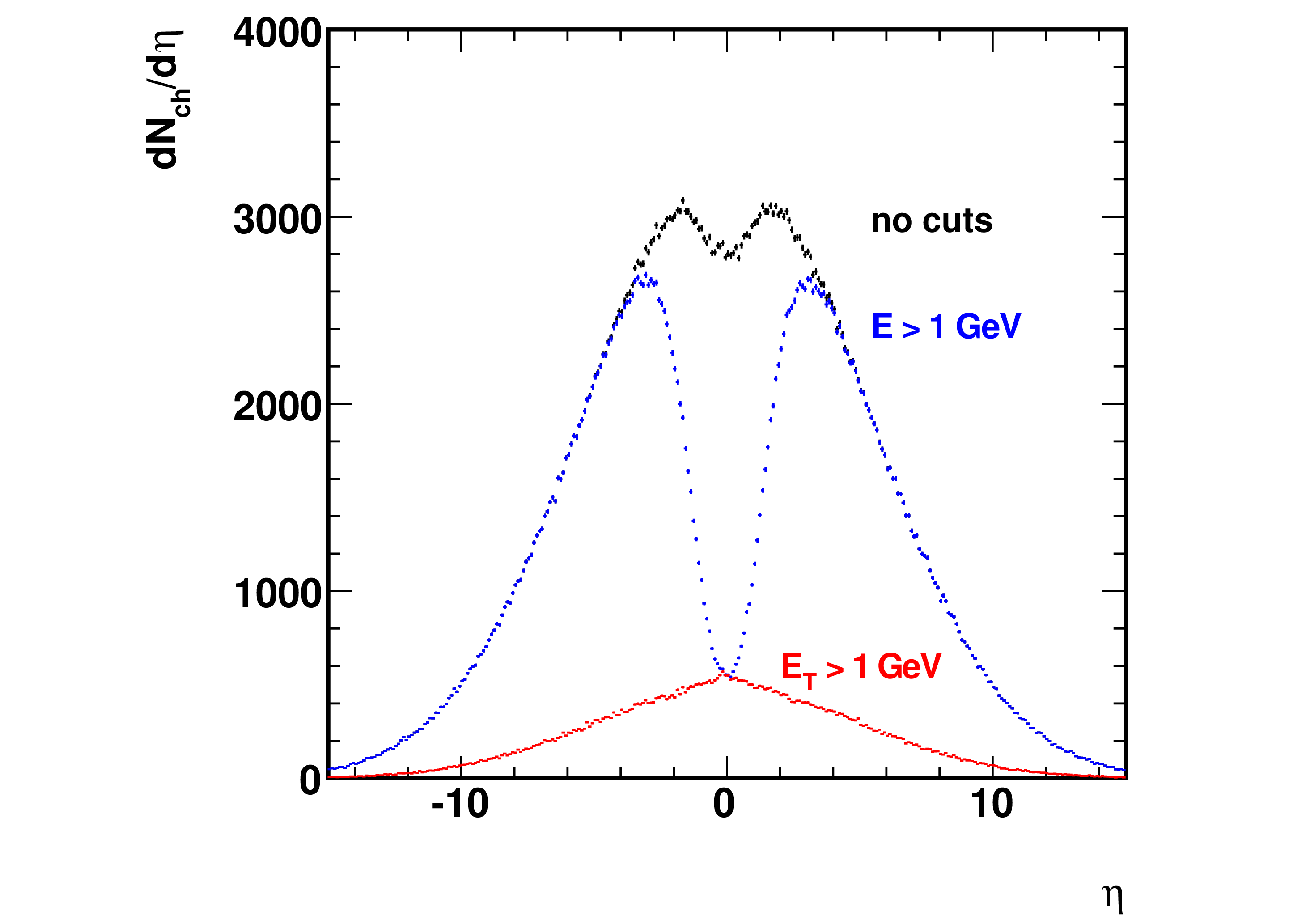}
 \end{minipage}

 \caption{Left: Estimated loss of low multiplicity events due to triggering
requirements on nTowers for cuts on $E$ in minimum bias p-p collisions.
Right: Pseudo-rapidity distribution of charged hadrons in central Pb-Pb
collisions at 5.5~TeV from the Hydjet generator. Particle selection to mimic
the level-1 trigger is applied for total $\langle E \rangle$ and transverse
$\langle E_T \rangle$ energy.}

 \label{fig:hftrig}
\end{figure}

The minimum bias trigger will be based on the requirement of a symmetric
number of hits in both forward calorimeters ($3<|\eta|<5$, see
Fig.~\ref{fig:hftrig}). For Pb-Pb collisions the centrality
trigger will be provided by correlating barrel and forward energies. The
charged particle multiplicity can be measured event-by-event using hits in
the innermost pixel layer with about 2\% accuracy and systematics below 10\%.

%\subsection{Low $p_T$ hadron spectra}

\begin{figure}[!h]
 \begin{center}
 \includegraphics[width=0.37\textwidth]{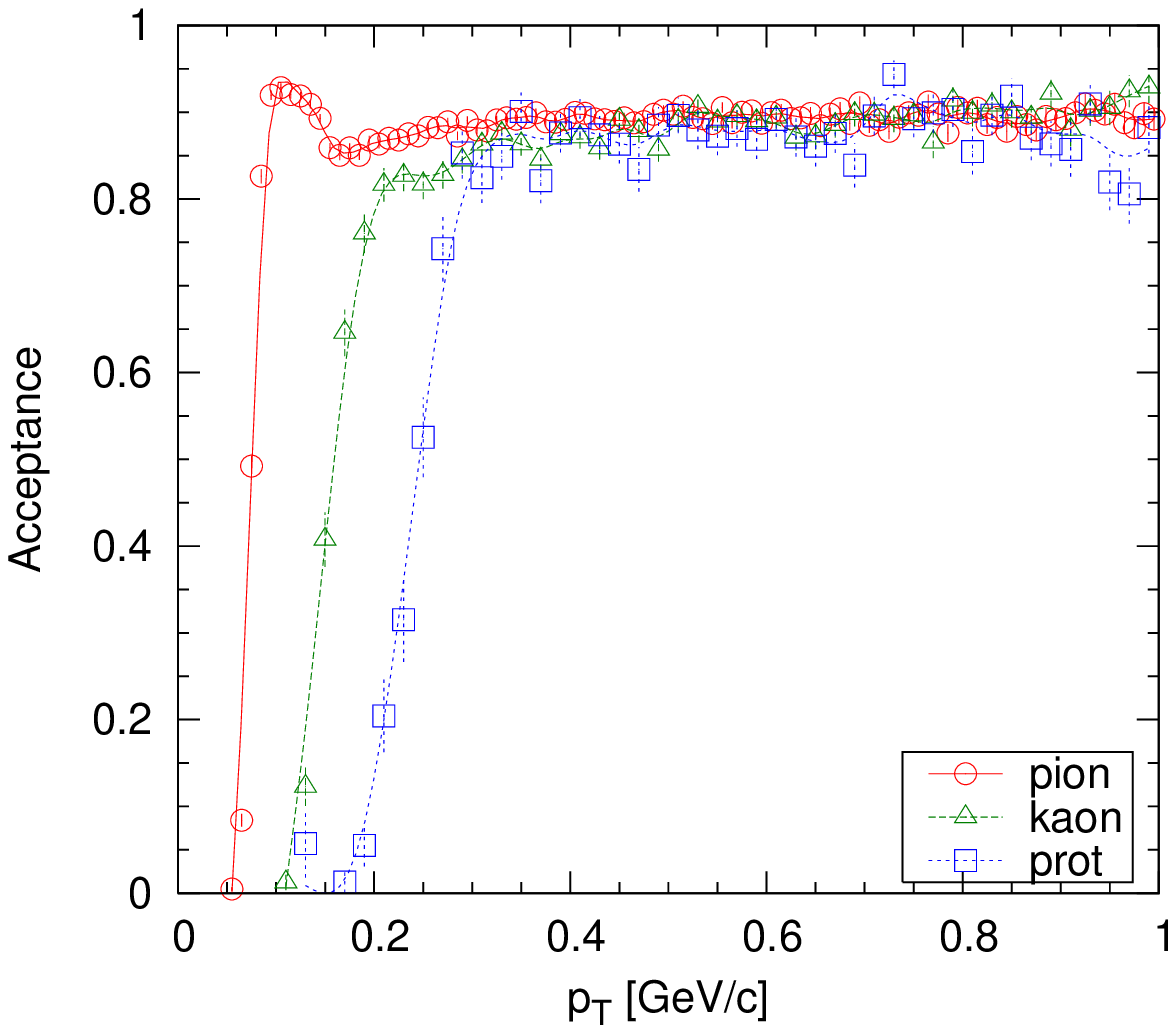}
 \includegraphics[width=0.37\textwidth]{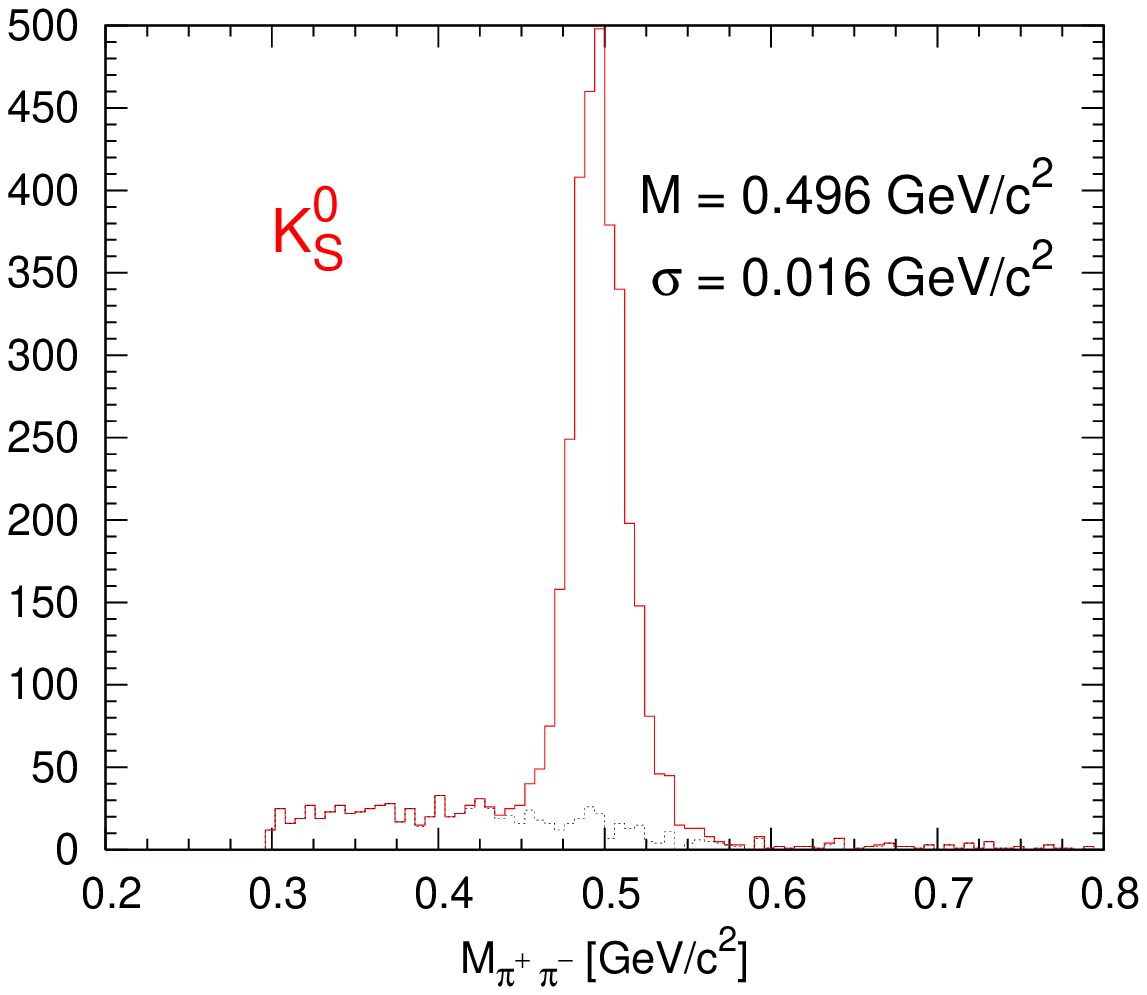}
 \includegraphics[width=0.24\textwidth]{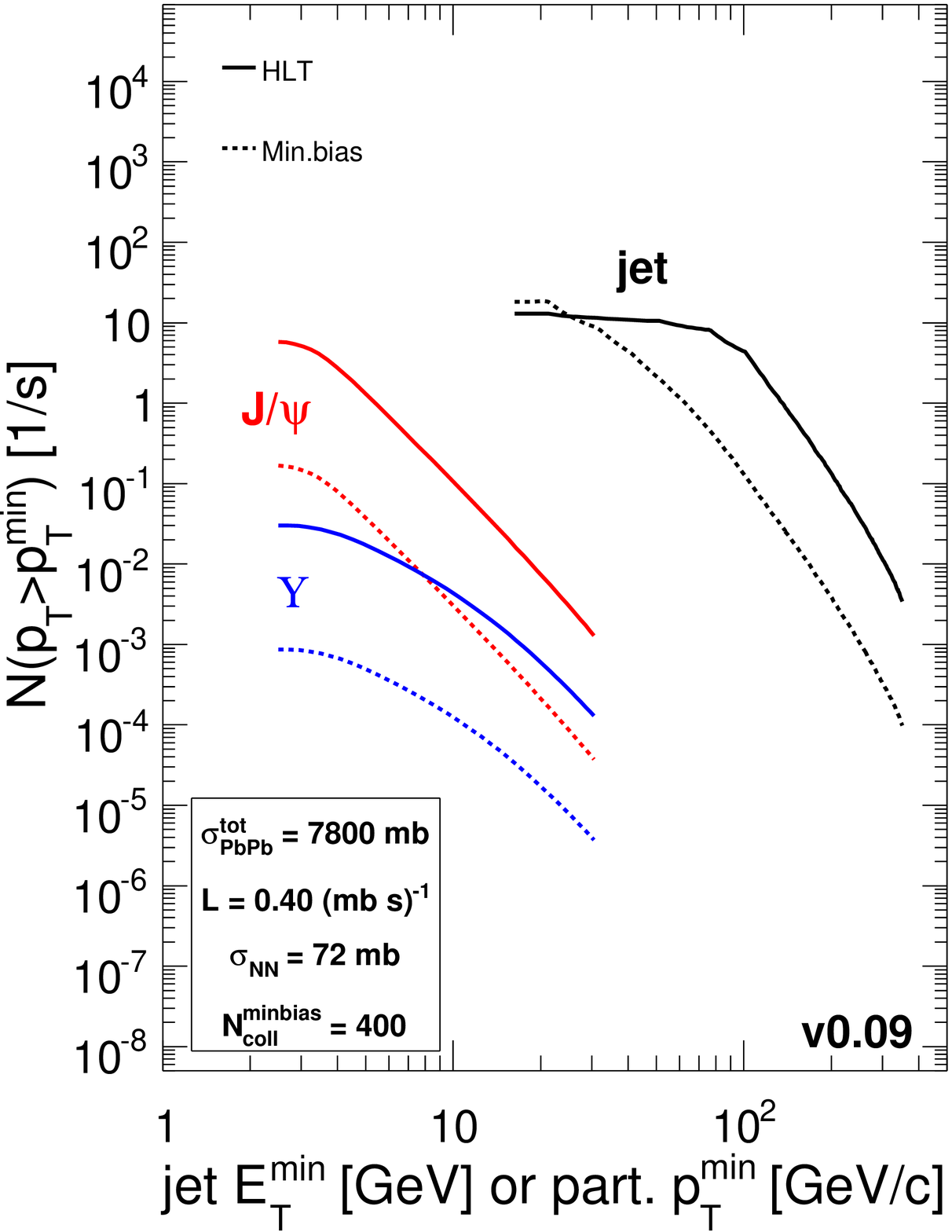}
 \end{center}

 \caption{Left: Acceptance of the track reconstruction algorithm as a
function of $p_T$, for tracks in the range $|\eta|<1$. Values are given
separately for pions (circles), kaons (triangles) and (anti)protons
(squares). Center: Invariant mass distribution of reconstructed
$\mathrm{K^0_S}\rightarrow \pi^+\pi^-$ in single minimum bias p-p collisions.
The mass distribution of the background is indicated with a black dashed
histogram.  Right: Minimum bias and high level trigger $\mathrm{J/\psi}$,
$\Upsilon$, and jet trigger rates for design luminosity in central Pb-Pb
collisions.}

 \label{fig:lowpt}
\end{figure}

CMS can study soft physics better than previously thought. Using a
modified pixel hit triplet finding algorithm, charged particles down to very low $p_T$
can be reconstructed (Fig.~\ref{fig:lowpt}-left). Particle identification
using energy loss in silicon is possible if $p<1-2$ GeV/$c$, benefitting
from analogue readout. Acceptances and efficiencies are at 80--90\%, the $p_T$
resolution is about 6\%.  At the same time low fake track rate is achieved
thanks to the geometrical shape of the hit cluster: below
10\% even in central Pb-Pb for $p_T>0.4$ GeV/$c$. This enables the study of identified particle
spectra (down to $p_T$ of $0.1-0.3$ GeV/$c$) and yields, multiplicity
distributions and correlations. Weakly decaying resonances are
accessible if the found tracks are combined and selected via decay topology:
strange neutral particles ($\mathrm{K^0_S}$, Fig.~\ref{fig:lowpt}-center,
$\Lambda$, $\overline{\Lambda}$), multi-strange baryons ($\Xi^-$,
$\Omega^-$). Also open charm ($\mathrm{D^0}$, $\mathrm{D^{*+}}$) and open beauty
($\mathrm{B} \rightarrow \mathrm{J/\psi} + \mathrm{K}$) can be studied.

%\subsection{Elliptic flow}

In Pb-Pb collisions azimuthal correlations give information on the
viscosity and parton density of the produced matter. The event plane can be
reconstructed using calorimetry.  The estimated event plane resolution is
about 0.37~rad if $b =$ 9~fm. The second moment $v_2$ can be measured with
about 70\% accuracy.  The results will improve by adding tracker information
and using forward detectors, such as the zero degree calorimeter.

%\subsection{High level trigger}

\section{Hard physics}

Interesting events are selected first by the level-1 trigger. It is a fast
hardware trigger, decisions are made within about 3~$\mu$s after the
collision. It mostly uses signals from the muon chambers and calorimeters.
After that step the event rate is still high, the efficient observation of
rare hard probes requires a high level trigger (HLT).  The trigger uses about
ten thousand CPUs working with the full event information including data from
the silicon tracker. A detailed study has been done with running offline
algorithms by parametrising their performance. Trigger tables are produced
considering various channels and luminosity scenarios
(Fig.~\ref{fig:lowpt}-right).

\begin{figure}
 \begin{center}
  \includegraphics[width=0.49\textwidth]{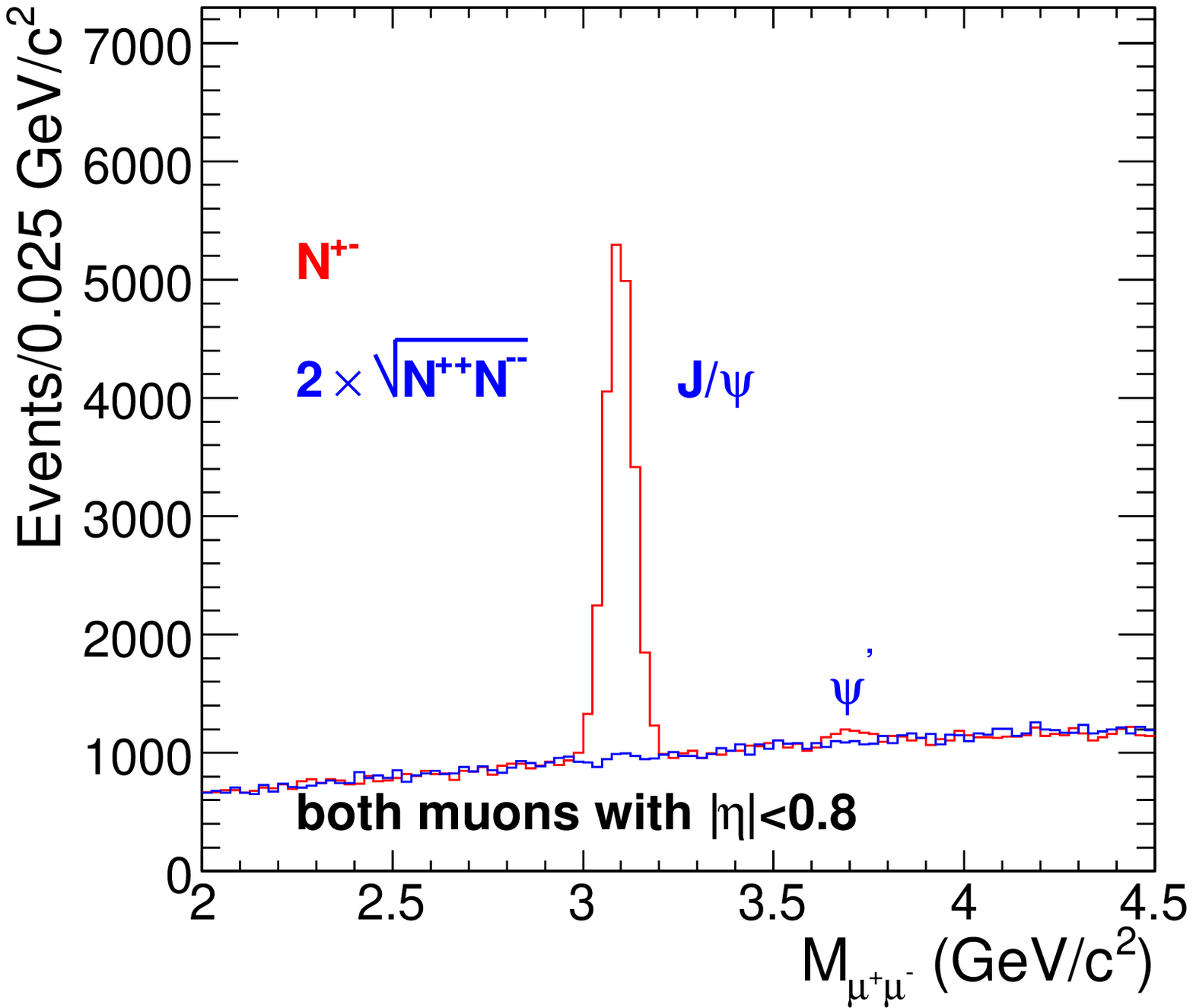}
  \includegraphics[width=0.49\textwidth]{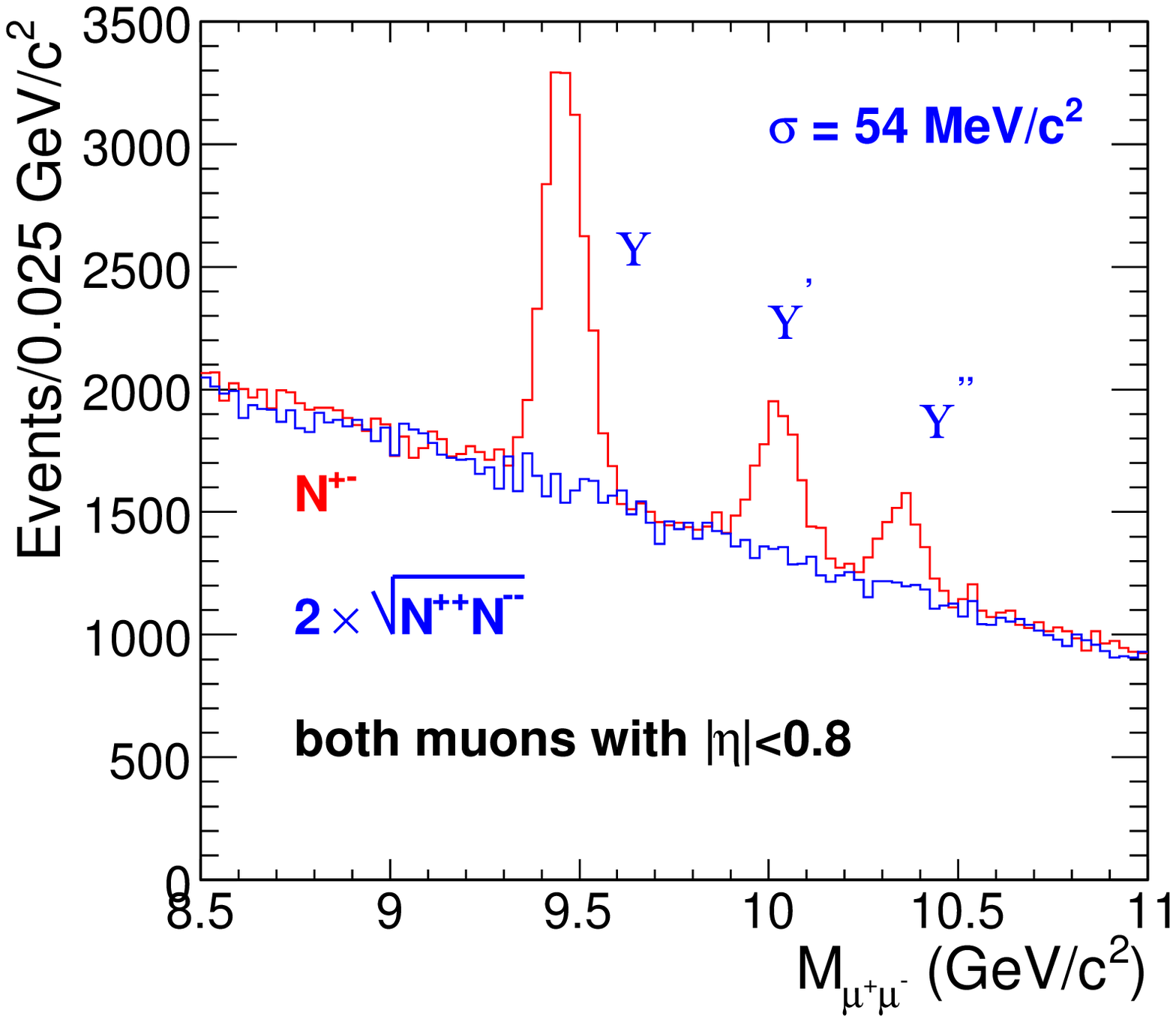}
 \end{center}
 \caption{(color online) Invariant mass spectra of opposite-sign and like-sign muon pairs
with $\mathrm{d}N_{ch}/\mathrm{d}\eta|_{\eta=0}$ = 2500, in the
$\mathrm{J/\psi}$ (left) and $\Upsilon$ (right) mass regions.}
 \label{fig:qqbar}
\end{figure}

%\subsection{Quarkonia}

Charmonium and bottomonium resonances can report on the thermodynamical state
of the medium via their melting. It is an open question whether they are
regenerated or suppressed at LHC energy. They can be reconstructed in the
dimuon decay channel with help of precise tracking. Acceptances are at 25\%
($\Upsilon$) and 1.2\% ($\mathrm{J/\psi}$) with 80\% efficiency and 90\%
purity. The mass resolution is 86~MeV/$c^2$ at the $\Upsilon$ mass and
35~MeV/$c^2$ at the $\mathrm{J/\psi}$ mass, in the full acceptance,
and even better in the barrel
(Fig.~\ref{fig:qqbar}). This is the best resolution achieved at the LHC. With
help of the HLT, 50 times more $\mathrm{J/\psi}$ and 10 times more $\Upsilon$ will be
collected.

%\subsection{Jets}

\begin{figure}[h]
 \begin{center}
  \includegraphics[width=0.45\textwidth]{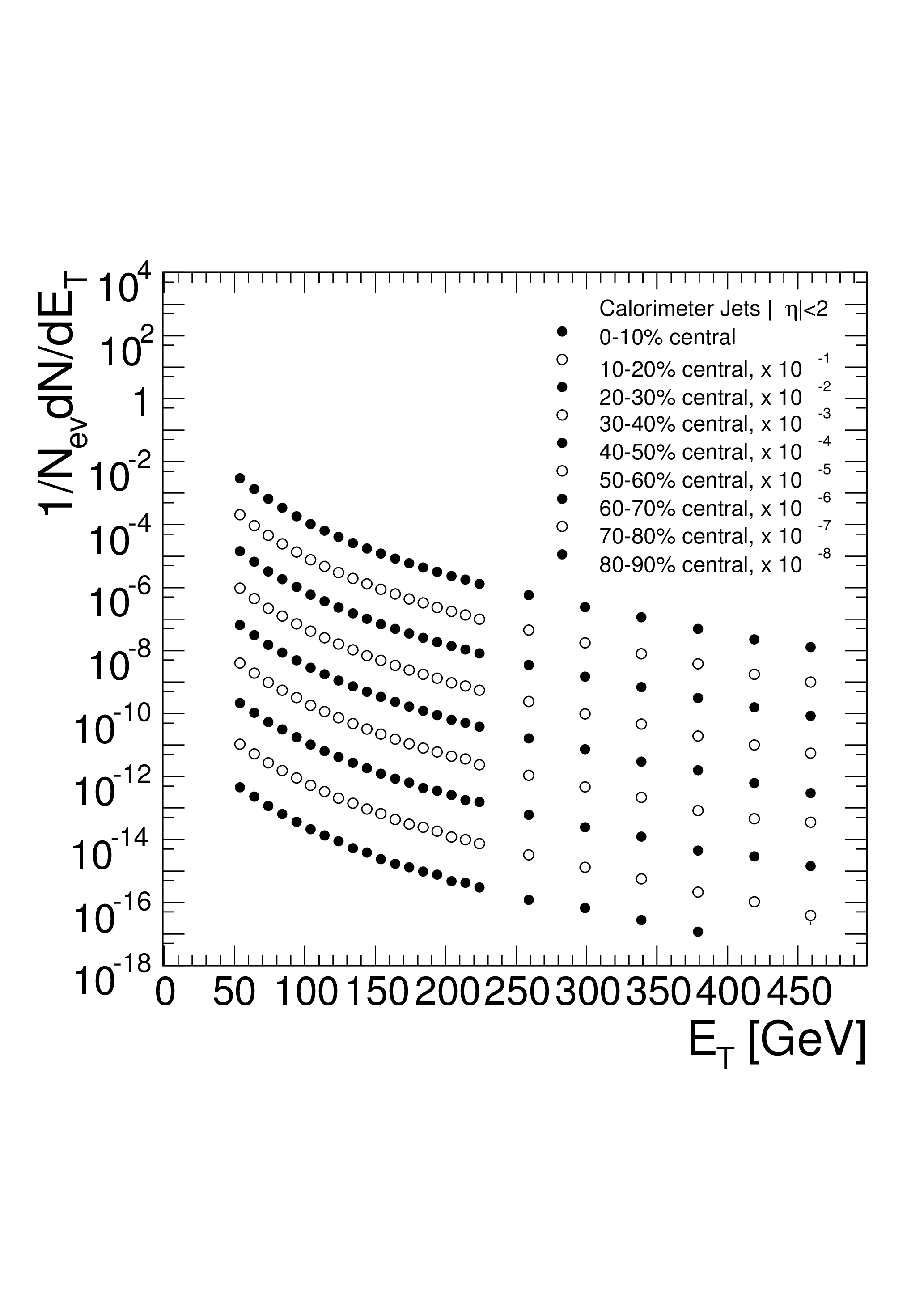}
  \hspace{0.05\textwidth}
  \includegraphics[width=0.43\textwidth]{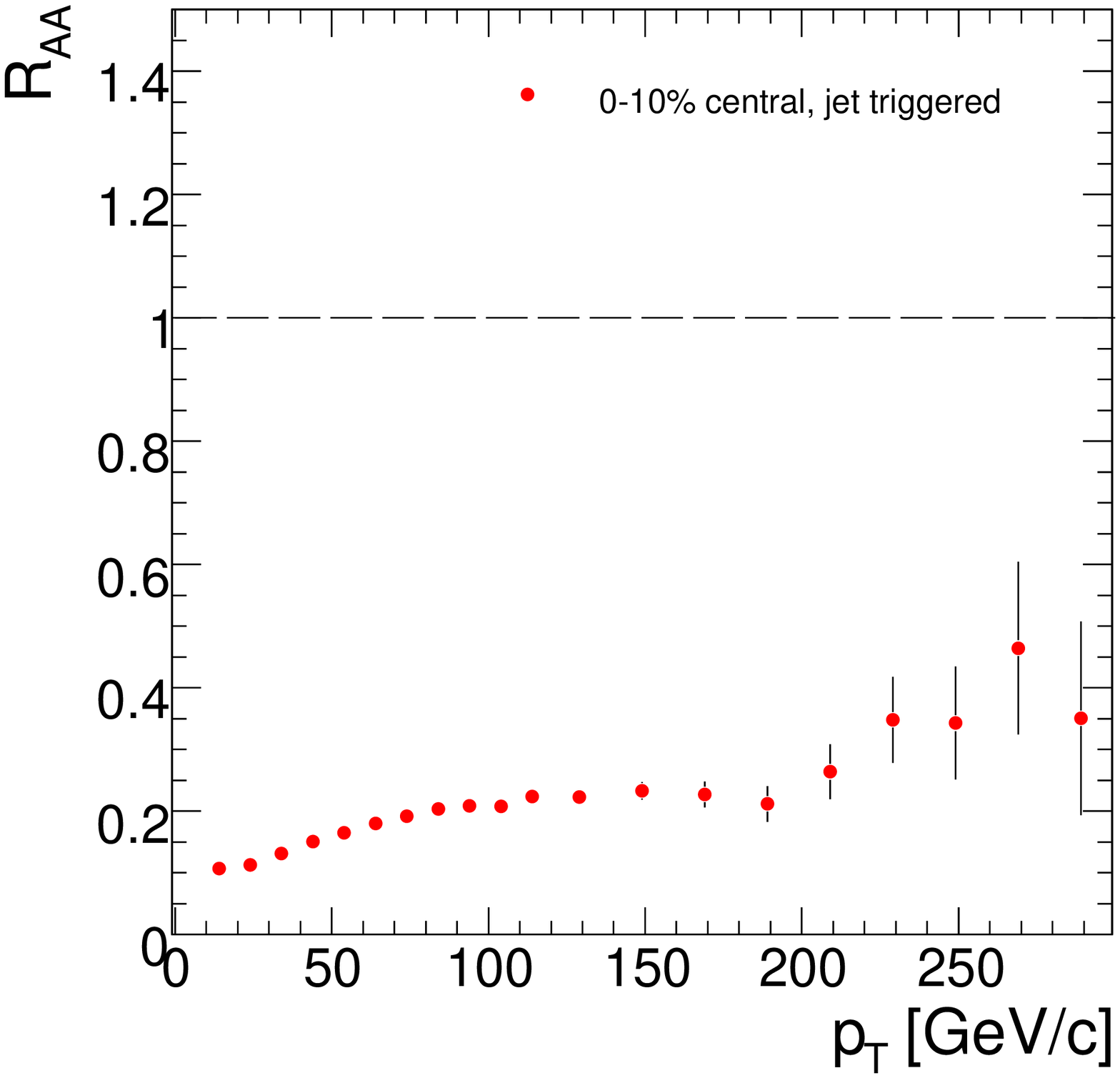}
 \end{center}

 \caption{Left: Expected inclusive jet $E_T$ distributions in 10 centrality
bins. Right: Expected statistical reach for the nuclear modification factor
for inclusive charged hadrons. For both figures, central Pb-Pb collisions at
5.5~TeV have been generated by Hydjet, with integrated luminosity of
0.5~$\mathrm{nb^{-1}}$.
%Triggered data sets are merged by the scaling procedure.
}
 \label{fig:hltjet}
\end{figure}

Finding jets on top of a high background is a challenge in Pb-Pb collisions.
Jets are reconstructed using a pile-up subtraction algorithm. It consists of
an
iterative jet cone finder and an event-by-event background subtraction. For
100~GeV jets the directional resolutions are $\sigma_\eta \approx$ 2.8\%,
$\sigma_\phi \approx$ 3.2\%, while the energy resolution is $\sigma_{E_T}
\approx$ 16\%. Thanks to the HLT, the reach of the jet $E_T$ measurement can
be
extended to about 0.5~TeV (Fig.~\ref{fig:hltjet}-left). The data sets,
triggered with 50, 75 and 100 GeV, are merged with a simple scaling
procedure.

%\subsection{High $p_T$ hadrons}

Parton energy loss in the hot and dense medium created in Pb-Pb collisions
can be studied by measuring the nuclear modification factors $R_{AA}$ and
$R_{CP}$. High $p_T$ charged particles can be tracked with about 75\%
algorithmic efficiency, few percent fake track rate for $p_T>1$ GeV/$c$ and
excellent momentum resolution. Using the HLT, the $p_T$ reach of the measurement is
extended from 90 to 300~GeV/$c$ (Fig.~\ref{fig:hltjet}-right).
%This allows for precise differential studies of high $p_T$ suppression.

\section{Forward physics}

%\subsection{Ultraperipheral collisions}

\begin{figure}[h]
 \begin{center}
 \begin{minipage}[c]{0.29\textwidth}
  \includegraphics[width=\textwidth]{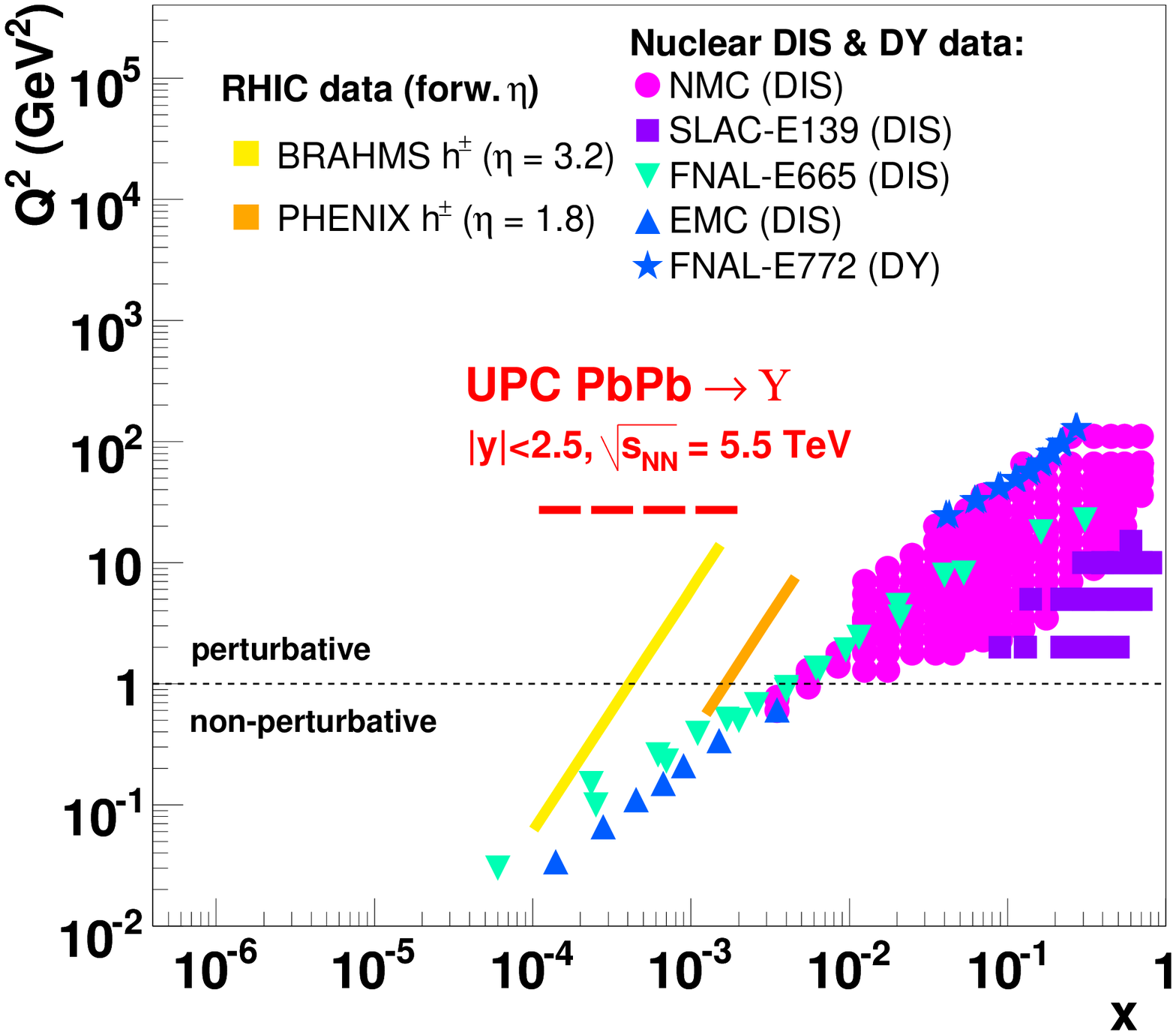}
 \end{minipage}
 \hspace{0.02\textwidth}
 \begin{minipage}[c]{0.33\textwidth}
  \includegraphics[width=\textwidth]{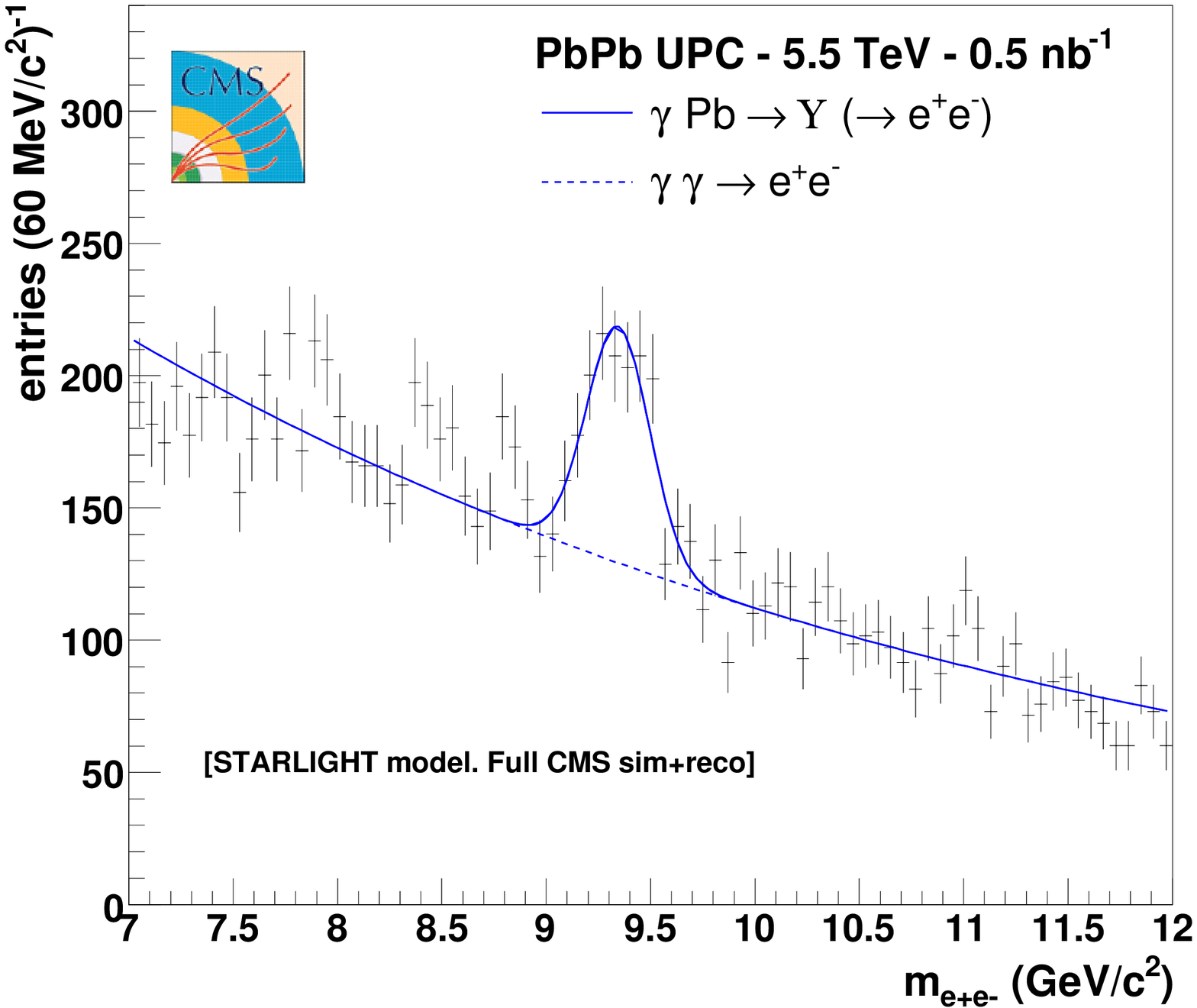}
 \end{minipage}
 \begin{minipage}[c]{0.33\textwidth}
  \includegraphics[width=\textwidth]{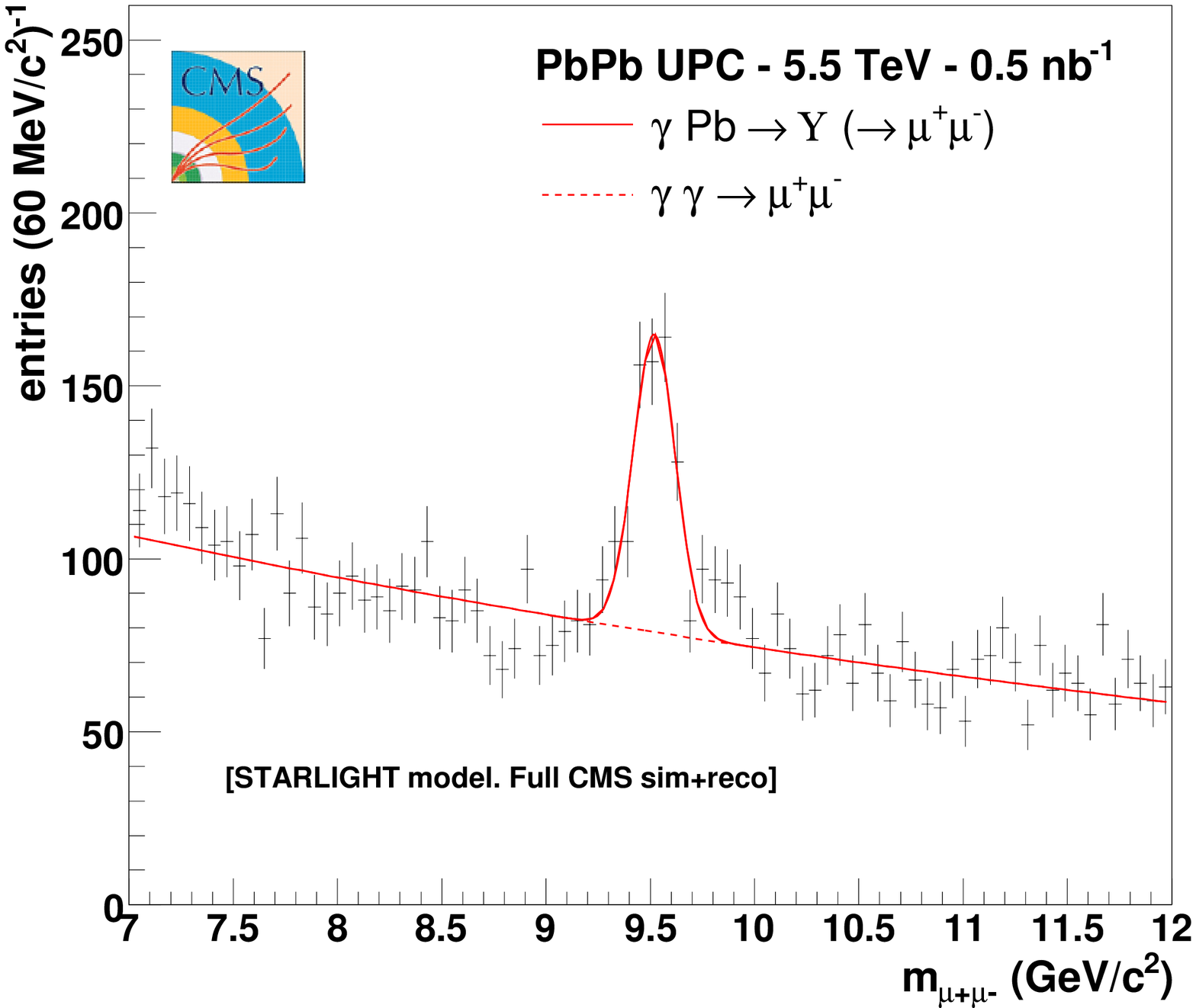}
 \end{minipage}
 \end{center}
 \caption{Left: The approximate $(x,Q^2)$ range covered by photoproduction in
ultraperipheral Pb-Pb collisions at the LHC is indicated. Right: Invariant
mass $\mathrm{e^+e^-}$ and $\mu^+\mu^-$ distributions for photoproduced
$\Upsilon$ and dilepton continuum, as expected in ultraperipheral Pb-Pb
collisions at 5.5~TeV, for integrated luminosity of 0.5~$\mathrm{nb^{-1}}$.}
 \label{fig:upc}
\end{figure}

The study of diffractive photoproduction of vector mesons in ultraperipheral
Pb-Pb collisions can constrain the gluon density at small $x$
(Fig.~\ref{fig:upc}-left). The decay channels $\rho\rightarrow\pi^+\pi^-$ and
$\Upsilon\rightarrow\mathrm{e}^+\mathrm{e}^-$ or $\mu^+\mu^-$ have been
studied, tagged with forward neutron detection in the zero degree
calorimeter. The combined acceptance and efficiency of the method is around 20\%
and it gives a good mass resolution in both channels
(Fig.~\ref{fig:upc}-centre and right).
%The expected yields of $\Upsilon$ mesons enables detailed studies of $p_T$
%and rapidity dependence.

\section{Summary}

The CMS detector combines capabilities for global event characterization and
for physics with specific probes. It performs equally well in soft, hard and
in forward physics, often supported by high level triggering.

\section*{Acknowledgment}

The author wishes to thank to the Hungarian Scientific Research Fund (T
048898).

\end{document}